\documentclass[12pt]{article}
\begin{document}
\title{Implicit Solutions of  PDE's}\author{ D.B. 
 Fairlie$\footnote{e-mail: david.fairlie@durham.ac.uk}$\\
\\
Department of Mathematical Sciences,\\
         Science Laboratories,\\
         University of Durham,\\
         Durham, DH1 3LE, England}
\maketitle

\begin{abstract}
Further investigations of implicit solutions to non-linear partial differential equations are pursued. Of particular interest are the equations which are Lorentz invariant. The question of which differential equations of second order for a single unknown $\phi$ are solved by the imposition of an inhomogeneous quadratic
relationship among the independent variables, whose coefficients are functions of $\phi$ is discussed, and it is shown that if the discriminant of the quadratic 
vanishes, then an implicit solution of the so-called Universal Field Equation is obtained. The relation to the general solution is discussed.
\end{abstract}

\section*{Implicit Solutions of  PDE's}
\section{Introduction}
The study of nonlinear equations particularly those incorporating Lorentz invariance  has demonstrated that a characteristic feature of large classes of solutions, and sometimes  the generic solution is that solutions are obtainable only in implicit form. As such, a few implicit solutions to nonlinear equations are recorded in the literature,
\cite{chaundy}\cite{gov}\cite{dbf2}\cite{dbf3}\cite{dbflez}\cite{thom1}
\cite{thom2}
, one of the neatest being the result that the general solution of the
Bateman equation,
\[(\phi_x)^2\phi_{tt} - 2\phi_x\phi_t\phi_{tx} + (\phi_t)^2\phi_{xx}\,=\,0,\]
where subscripts denote derivatives , 
 is given by the implicit solution of the following constraint between two arbitrary functions of one variable;
\begin{equation}  tF(\phi) + xG(\phi) \,=\,1,\label{sol2}\end{equation}
where  $F,\ G$ are arbitrary functions of $\phi$. Since this solution of a second order partial differential equation depends upon two arbitrary functions of the space variables for a hypebolic equation, and the functional form of $(F,\ G)$ an be fitted to the intitial configuration and time deivative, it is the most general. 
This result is well known. One of the simplest ways of understanding it
is to look for a solution in the implicit form
\[t\,=\,A(\phi ,x),\]
Differentiation of this equation with respect to $(t,\ x)$ and substitution of 
the partial derivatives of $\phi$ into the Bateman equation shows that this function will be a solution provided 
$\displaystyle{\frac{\partial^2 A(\phi ,x)}{\partial x^2}\,=0}$.
But this simply means that $A(\phi, x)$ is linear in $x$, i.e.
\[ t\,=\, A_0(\phi) +A_1(\phi)x,\]
which is tantamount to (\ref{sol2}).
Later we shall generalise this approach.
The Bateman equation possesses a remarkable property of  covariance, which is reflected in the transformation used to obtain it, and also in form of the solution (\ref{sol2}), namely that if $\phi(x,t)$ is a solution. so is any function of  $\phi$. 
In this paper, instead of asking for the solution to a given nonlinear PDE, we ask for those equations which are solved by an implicit functional relationship among the
unknown and the independent variables
A natural attempt to generalize the Bateman result to the linear constraint
\[\sum_{i=1}^{i=n}F_i(\phi)x_i\,=\,{\rm constant},\]
leads to the surprising conclusion that the function $\phi$ defined thereby  is a `universal' solution to any equation of motion arising from any Lagrangian dependent upon $\phi$ and its first derivatives, and homogeneous of weight one in the latter \cite{dbfeuler}.
It so happens that if there are only two independent variables, the  equation  resulting
from this prescription is unique, and is the Bateman equation. In the case of more than two independent variables, however, the linear constraint cannot be designed to fit arbitrary initial conditions and is not the most general solution.
One of the  aims of this paper is to extend this study to quadratic constraints of the form
\begin{equation}
\sum_{i=1}^{i=n}\sum_{j=1}^{j=n}M_{ij}(\phi)x_ix_j\,=\,1.\label{quad}
\end{equation}
Note that both the linear and quadratic constraints remain form-invariant under a general linear transformation of the independent variables $x_i$, hence under Lorentz or Euclidean transformations, so viewed as classes of implicit solutions, they should apply to Lorentzinvariant equiations.
The quadratic system has been studied before, by M. Dunajski \cite{dun}, from a different point of view. He proposes to determine the conditions upon $\phi$ such that an equation of D'Alembertian type;
\[\frac{\partial}{\partial x_i}\left(\eta_{ij}\frac{\partial\phi}{\partial x_j}\right)\,=\,0\]
admits a solution on the manifold defined by  (\ref{quad}). A particular example of this occurs as an exercise in Jeans' famous book on Electromagnetism \cite{jeans}
`Shew that the confocal ellipsoids
\[\frac{x^2}{a^2+\lambda}+ \frac{y^2}{b^2+\lambda}+ \frac{z^2}{c^2+\lambda}\,=\,1\] 
can form a system of equipotentials and express the potential as a function of $\lambda.'$
If we take  (\ref{quad}) and differentiate it with respect to $x_p$ the result is
\begin{equation}
\sum_{i=1}^{i=n}\sum_{j=1}^{j=n}M'_{ij}(\phi)x_ix_j\phi_p+\sum_{j=1}^{j=n} 2M_{pj}(\phi)x_j \,=\,0\label{quad2}.
\end{equation}
Here $\phi_p$ denotes differentiation of $\phi$ with respect to $x_p$, and a prime
means differentiation with respect to the argument.
Differentiation again gives
\begin{eqnarray}
&&M''_{ij}(\phi)x_ix_j\phi_{p}\phi_{q}+ M'_{ij}(\phi)x_ix_j\phi_{pq}\nonumber\\
&&+2\left(M'_{qj}(\phi)x_j\phi_p+M'_{pj}(\phi)x_j\phi_q+M_{pq}(\phi)\right)\,=\,0, \label{quad3}
\end{eqnarray}
where repeated indices imply summation.
Multiplication by $\eta_{pq}$,  summing  and substituting from (\ref{quad2}) for $\phi_p$ etc and integrating gives the equation (2.9) of  Dunajski's paper 
\[(gM'_{ij}-M'^{ik}\eta_{km}M'^{mj})x_ix_j\,=\,0.\]
where $2g'\,=\, \eta_{ij}M^{ij}$.
This equation is the starting point to his analysis.
\section{Alternative interpretation.}
Here the purpose  is to ask the different question `What differential equations 
possess  solutions of implicit form given by the quadratic ansatz (\ref{quad})?'
This is an obvious extension of the linear situation.
Taking equations (\ref{quad3}) for indices $(pp,pq,qq)$ respectively it is possible to
eliminate the second derivatives of $M_{ij}$ and at the same time many of the other terms. If we define $\lambda$ by
\[2\lambda\,=\,-\sum_{i=1}^{i=n}\sum_{j=1}^{j=n}M'_{ij}(\phi)x_ix_j\]
and $f_{pq}$ by
\begin{equation}
f_{pq} =(\phi_p)^2\phi_{qq} - 2\phi_p\phi_q\phi_{pq} + (\phi_q)^2\phi_{pp}\label{def}
\end{equation}
where subscripts denote partial differentiation with respect to $x_p$ etc,
so that $f_{pq}\,=\,0,$ (\ref{quad5}) is simply the Bateman equation in the independent variables
$(x_p,\ x_q)$, the resultant eliminant is just
\begin{equation}
\lambda f_{pq}  + \left(M_{pp}(\phi_q)^2-2M_{pq}\phi_p\phi_q+M_{qq}(\phi_p)^2\right) \,=\,0.\label{quad4}
\end{equation}
In fact, there is a more general eliminant, namely
\begin{eqnarray}
&&(\lambda \phi_{pq} + M_{pq})\phi_r\phi_s -(\lambda \phi_{rq} + M_{rq})\phi_p\phi_s\nonumber\\
&-&(\lambda \phi_{ps} + M_{ps})\phi_r\phi_q +(\lambda \phi_{rs} + M_{rs})\phi_p\phi_s
\,=\,0\label{gen}
\end{eqnarray}
Rewriting (\ref{quad2}) in the form
\begin{equation}
\lambda\phi_p + M_{pp}x_p+\sum_{j\neq p}M_{pj}x_j\,=\,0\label{quad5}
\end{equation}
The off diagonal elements of the matrix may then be eliminated from (\ref{quad5}) by using (\ref{quad4}). We see that (\ref{quad4}) and (\ref{quad5}) constitute $\frac{1}{2}n(n+1)$ linear equations for the elements of the symmetric matrix $M$. Imposition of one further relationship upon the matrix elements, which may be taken as homogeneous and nonlinear, will leave a nonlinear equation of second order for $\phi$. In the next section this procedure is carried out in detail for two independent variables.  
\section{Two variable case}
In the case of two independent variables the relevant equations are
\begin{eqnarray}
\lambda\phi_1 + M_{11}x_1+M_{12}x_2&=&0,\nonumber\\
\lambda\phi_2 + M_{22}x_2+M_{12}x_1&=&0,\nonumber\\
\lambda f_{12}  + \left(M_{11}(\phi_2)^2-2M_{12}\phi_1\phi_2+M_{22}(\phi_1)^2\right) &=&0.\nonumber
\end{eqnarray}
The easiest way to manage this is to solve these equations for \hfill\break
$M_{11},\ M_{12},\ M_{22}$, giving
\begin{eqnarray}
M_{11}&=&-\frac{\lambda(x_1\phi_1^3+x_2\phi_1^2\phi_2+x_2^2f_{12})}
{(x_1^2\phi_1^2+2x_1x_2\phi_1\phi_2+x_2^2\phi_2^2)}\,=\,\ \ \ \lambda^2(\phi_1^2- \lambda x_2^2f_{12}),\nonumber\\
M_{22}&=&-\frac{\lambda(x_1^2f_{12}+x_1\phi_1\phi_2^2+x_2\phi_2^3)}
{(x_1^2\phi_1^2+2x_1x_2\phi_1\phi_2+x_2^2\phi_2^2)}\,=\,\ \ \ \lambda^2(\phi_2^2 -\lambda x_1^2f_{12}),\nonumber\\
M_{12}&=&-\frac{\lambda(x_1\phi_1^2\phi_2+x_2\phi_1\phi_2^2-x_1x_2f_{12})}
{(x_1^2\phi_1^2+2x_1x_2\phi_1\phi_2+x_2^2\phi_2^2)}\,=\,\lambda^2(\phi_1\phi_2+ \lambda x_1x_2f_{12}).\nonumber
\end{eqnarray}
\begin{itemize}
\item{Example 1.}
The imposition of the homogeneous relationship
\[M_{11}M_{22}-M_{12}^2 \,=\, 0,\]
 yields the differential equation
\[{f_{12}}\,=\,0.\]

That this is just the Bateman equation is to be expected, since the
relationship imposed makes the initial quadratic factorize into the square of (\ref{sol2}). 
\item{Example 2.}
Imposition of the constraint $M_{12}\,=\,0$ which unlike the previous constraint breaks the Lorentz invariance, unless $(x_1,\ x_2)$ are considered as light cone co-ordinates yields the equation
\begin{equation}
\phi_1^2\phi_2x_1 + \phi_1\phi_2^2x_2- x_1x_2f_{12} \,=\,0.\label{ex}
\end{equation}
Now the constraint equation (\ref{quad}) becomes 
\[M_{11}(\phi)x_1^2+M_{22}(\phi)x_2^2 \,=\,1,\]
which gives the general solution for the Bateman equation expressed in the variables $(x_1^2,\ x_2^2)$, which becomes  the equation (\ref{ex}) when written  in the original variables $(x_1,\ x_2).$
\end{itemize}
A similar result can be found in the three variable case; If the discriminant of the quadratic form is set to zero, i.e.

\begin{equation}
\det\left|\begin{array}{ccc}
               M_{11}&M_{12}&M_{13}\\
               M_{12}&M_{22}&M_{23}\\
               M_{13}&M_{23}&M_{33}\end{array}\right|=0,
\label{disc}
\end{equation}
which remains compatible with the form-invariance of the ansatz under linear co-ordinate transformations,
then the resulting equation is what has been called by its authors `the universal field equation',\cite{gov}, because it arises from an infinite number of inequivalent Lagrangians, 
\begin{equation}
\det\left|\begin{array}{cccc}0&\phi_1&\phi_2&\phi_3\\
               \phi_1&\phi_{11}&\phi_{12}&\phi_{13}\\
               \phi_2&\phi_{12}&\phi_{22}&\phi_{23}\\
               \phi_3&\phi_{13}&\phi_{23}&\phi_{33}\end{array}\right|=0.
\label{univ}
\end{equation}
 This result is a special case, shortly to be generalized.
 \subsection{Multivariable case}

The simplicity of the solution suggests that it might be possible to solve equations (\ref{quad4}) and (\ref{quad5}) in general. Indeed this is so, and the solution is,
using the result responsible for the simplification above, namely
\[\sum_j\lambda\phi_jx_j\,=\,-1.\]
The result is
\begin{eqnarray}
&&M_{pq}\,=\,\nonumber\\
&&\lambda^2\left(\phi_p\phi_q-\frac{1}{2}\left((\sum_{r\neq p,q}\phi_r x_r)\frac{f_{pq}}{\phi_p\phi_q}-\phi_p\sum_{r\neq p}\frac{f_{qr}x_r}{\phi_q\phi_r}-\phi_q\sum_{r\neq q}\frac{f_{pr}x_r}{\phi_p\phi_r}-\phi_p\phi_q\mu\right)\right),\nonumber
\end{eqnarray}
where
\[\mu\,=\, \lambda\sum_{r,s}\frac{f_{rs}x_rx_s}{\phi_r\phi_s}.\]
In particular;
\begin{equation}
M_{pp} \,=\,\lambda^2\left(\phi_p^2+\sum\frac{x_r}{\phi_r}f_{pr}-\phi_p^2\mu\right).
\end{equation}
The imposition of any relation among the $M_{pq}$ 
replacing, if necessary, the parameter $\frac{1}{\lambda}$ by $-\sum\phi_ix_i$  will provide a second order differential equation whose solution is given in implicit form by the original quadratic relation (\ref{quad})
In particular, the imposition of the relation $\det|M_{jk}|\,=\,0$ imposes the condition that the implicit solution of the quadratic relation (\ref{quad}) for $\phi$
solves the Universal Field Equation
\begin{equation}
\det\left|\begin{array}{cc}0&\phi_i\\
                           \phi_j&\phi_{ij}\end{array}\right|\,=\,0\label{univ3}.\end{equation}
\underline{Proof}

Consider the equation
\begin{equation}
\det\left|\begin{array}{ccccc}0&\phi_1^2&\phi_2^2&\dots&\phi_n^2\\
                              \phi_1^2&0&f_{12}&\dots&f_{1n}\\
                              \phi_2^2&f_{12}&0&\dots&f_{2n}\\
                               \vdots&\vdots&\dots&\ddots&\vdots\\ 
                           \phi_n^2&f_{1n}&f_{2n}&\dots&0\end{array}\right|\,=\,0.\label{univ2}
\end{equation}
From the definition of $f_{pq}$, (\ref{def}), by elementary row and column operations
this equation is equivalent to
\[ (-1)^n 2^{n-1}\prod_{i=1}^{i=n}\phi_i^2\det\left|\begin{array}{cc}0&\phi_i\\
                           \phi_j&\phi_{ij}\end{array}\right|\,=\,0,\]
i.e. the Universal Field Equation.
On the other hand, from (\ref{quad4}) the left hand side of equation (\ref{univ2})
is equivalent to  
\begin{equation}
 (-1)^n \left(\frac{2}{\lambda}\right)^{n-1}\prod_{i=1}^{i=n}\phi_i^2\det\left|\begin{array}{cc}0&\phi_i\\
                           \phi_j&M_{ij}\end{array}\right|.\label{vanish}
\end{equation}
Now if $\det|M_{jk}|\,=\,0$, then for a consistent non trivial solution of the other $n$ linear equations which determine full set of the matrix elements $M_{pq}$ in addition to (\ref{quad4}), namely (\ref{quad5}), we see that every determinant obtained
from $\det|M_{jk}|$ by replacing each each column in turn by a column vector of the $\phi_i$, must vanish, hence the entire determinant (\ref{vanish}) vanishes, and 
the result is proved.

\section{Generalisation}
It it has already been observed that the Universal Field Equation is solved by a linear
assumption \cite{gov}\cite{dbfeuler}, and now it has been shown that a quadratic ansatz also works. In addition, explicit solutions can be obtained by taking $\phi$ as a homogeneous function of its arguments of weight zero \cite{gov}.   characterization of the general solution, which is known to be integrable, \cite{govlin} follows similar lines to those of \cite{dbflez}.

The clue as to how to do this comes from the discussion in the introduction in which solutions of the form 
\[t\,=\,A(\phi ,x_i),\ \, x_n =t,\] are sought. If this satisfies the Universal field equation, then the homogeneous Monge-Amp\`{e}re equation must be satisfied,
\[\det\left|\frac{\partial^2 A(\phi ,x_i)}{\partial x_i\partial x_j}\right|\,=\,0.\]
Both the linear and quadratic relations imposed here satisfy this constraint.
As a second example, consider the equation arising from the Lagrangian
$\sqrt{\phi_1^2+\phi_2^2+\phi_3^2}$, namely
\[f_{12}+f_{23}+f_{13}\,=\,0.\]
Setting $x_3 =t$, this requires as a condition upon $A$
\[(1+A_1^2)A_{22} +(1+A_2^2)A_{11}-2A_1A_2A_{12}\,,=\,0,\]
the so-called two dimensional Bateman equation, solution methods for which are described in \cite{nutku},\cite{mul}. Thus, as in the previous case, the general solution depends upon the implicit solution of a similar non-linear equation with one 
less independent variable.
\section{Complex Bateman equation}
This final chapter is a bit of a diversion from the main theme of the article, but is included as it is both anelegant result, obtained here in a slightly different way, and a nice example of how results in mathematical physics sometimes have been already anticipated several years earlier. 
The Bateman equation possesses a complex form 
\begin{equation}
\phi_x\phi_z\phi_{yw} - \phi_x\phi_w\phi_{yz} + \phi_y\phi_z\phi_{xw}-\phi_y\phi_w\phi_{xz}\,=\,0,\label{combat}\end{equation}
whose general solution has been published in 1935 by Chaundy \cite{chaundy} , rediscovered and generalized in 1999. \cite{leznov}.  
It is given by equating an arbitrary function of $(x,y,\phi)$ to another of $(z,w,\phi)$, 
i.e.
\[ F(x,y,\phi)\,\equiv\,G(z,w,\phi).\]
This equation may be expressed in first order form;
\begin{eqnarray}\frac{\partial u}{\partial x}&=& v\frac{\partial u}{\partial y},\label{one}\\
\frac{\partial v}{\partial z}&=& u\frac{\partial v}{\partial w},\label{two}\\
\frac{\partial \phi}{\partial x}&=& v\frac{\partial \phi}{\partial y},\label{three}\\
\frac{\partial \phi}{\partial z}&=& u\frac{\partial \phi}{\partial w}.\label{four}
\end{eqnarray}
The way this works is that we have
 \[ \phi_x \,=\, \frac{F_x}{G_\phi-F_\phi};\ \ \  \phi_y \,=\, \frac{F_x}{G_\phi-F_\phi}\]
\[ \phi_z\,=\, -\frac{G_z}{F_\phi-G_\phi};\ \ \  \phi_w \,=\, \frac{G_w}{F_\phi-G_\phi}\]
Hence, from (\ref{three}) and (\ref{four}) we have
\[ v\,=\,\frac {F_x}{F_y};\ \ \ u\,=\,\frac {G_z}{G_w}\]
i.e. 
\[ \frac{\partial u}{\partial x}\,=\,\left( \frac{G_{z\phi}}{G_w}-\frac{G_{w\phi}G_z}{G^2_w}
\right)\phi_x\,=\, \frac{\phi_x}{\phi_y}\frac{\partial u}{\partial y}\,=\,v\frac{\partial u}{\partial y},\]
thus verifying (\ref{one}), and similarly for (\ref{two}).

There is a huge class of explicit solutions to the equation (\ref{combat}).
Take any two arbitrary differentiable functions, $f(x,y)$ and $g(z,w)$ and construct an
arbitrary function $F(f(x,y),g(z,w))$. Then 
\[\phi\,=\,F(f(x,y),g(z,w))\]
is a solution to (\ref{combat}).
\section*{Acknowledgement}
I am indebted to Tatsuya Ueno for checking the calculations.

\end{document}